\documentclass[12pt]{article}
\usepackage {graphicx}
\def\beqa{\begin{eqnarray}}
\def\eeqa{\end{eqnarray}}
\def\beq{\begin{equation}}
\def\eeq{\end{equation}}

\begin{document}

\baselineskip=0.7truecm
\begin{titlepage}

\title{Some remarks on Bell's Inequality tests}

\author{Antonio Feoli\footnote{E-mail: feoli@unisannio.it}\\
  Dipartimento di Ingegneria, Universit\`{a} del
            Sannio\\ Corso Garibaldi 107, Palazzo Bosco Lucarelli \\ I-82100 Benevento, Italy \\
 INFN - Sezione di Napoli - Gruppo Collegato di Salerno, Italy\\ \\Salvatore Rampone\footnote{E-mail: rampone@unisannio.it}\\
DSGA, Universit\`a del Sannio,\\ Via Port'Arsa, 11 \\ I-82100
Benevento Italy\\ INFM - Unit\`a di Salerno, Italy}

 \maketitle

\begin{abstract}
We emphasize the difficulties of an experiment that can definitely
discriminate between local realistic hidden variables theories and
quantum mechanics using the Bell CHSH inequalities and a real
measurement apparatus. In particular we analyze some examples in
which the noise in real instruments can alter the experimental
results, and the nontrivial  problem to find a real ``fair sample" of
particles to test the inequalities.
\end{abstract}

\end{titlepage}
\section{Introduction}
In a classical paper~\cite{1}, Einstein, Podolsky, and Rosen
presented an argument which led them to infer that quantum mechanics
is not a complete theory. They believed  that the quantum mechanical
description of a physical system could be improved supplementing it
by new variables that can make quantum mechanics a local, realistic,
complete  and hopefully even deterministic theory. Postulating the
existence of hidden variables, Bell proved an inequality (discussed
in the next section) that must be satisfied by the local realistic
theories but that is not fulfilled by the statistical predictions of
quantum mechanics~\cite{3}. This inequality was generalized by
Clauser, Horne, Shimony, and Holt~\cite{4}, giving room to decisive
tests between Quantum Mechanics ($QM$) and Local Hidden Variable
($LHV$) theories.

After some contradictory results~\cite{e1}~\cite{e2}, between 1980
and 1982, Aspect et al.~\cite{a1}~\cite{a2} verified that in some
experiments the results were correctly predicted by quantum theory
and that in some configurations Bell's inequality was actually
violated. At our knowledge all experiments so far support quantum
theory~\cite{[5]}\cite{Bellexp}\cite{Tapster94}.

However these results have been the subject of several criticisms,
pointing out that there are ``experimental
loopholes"~\cite{[6]}~\cite{x} in Aspect's like experiments, divided
essentially into two main classes: ``locality" and ``detection
efficiency" loopholes, which might turn out to be responsible for the
result. For example Franson~\cite{crit} published a paper showing
that the timing constraint in one of these experiments was not
adequate to confirm that locality was violated. In a test with pairs
of entangled photons meeting polarizers, a good measurement apparatus
will rule out any possibility of subluminal communication which might
inform the photon or the detector on one side about which measurement
will be performed on the other side. For that purpose it has been
suggested that the orientation of the polarizers must be determined
by a purely random physical process and the photons must be
registered with accurate time tags separately at each detector before
any information can be communicated from the other side. Recently
Weihs et al.~\cite{viola} have finally claimed to have close the
locality loophole.

Other authors insist that all experiments so far have detected only a
small subset of all pairs created~\cite{[66]}, while Rowe et
al~\cite{Rowe} have recently performed an experiment without this
problem. Later we will briefly comment about this loophole.

Moreover the existence of some ``selection effect" which influences
the detection probabilities has been suspected, so that Aspect's
experiments would not actually test Bell's inequality~\cite{[7]}. So
far no experiment aiming at testing Bell's inequality seems to be
simultaneously free by all possible
loopholes~\cite{loophole}\cite{locality}\cite{closeloophole}\cite{Vaid}.

Recently, additional motivations to investigate quantum non-locality
have arisen, based on the potential applications of the fascinating
field of quantum information processing: all of quantum computation
and communication is based on the assumption that quantum systems can
be entangled and that the entanglement can be maintained over long
times and distances~\cite{PhysWorld}.

In this paper, starting from an ideal apparatus description, we
review and emphasize some potential critical points of a real
measurement system, in an information theory fashion. In particular
we propose some examples in which the noise plays a role in Bell's
inequality tests and we consider the selection effect due to the
choice of the sample of N particle pairs, independently of the
detection efficiency.

\section{Bell's Inequality}
Let us consider an ensemble of correlated pairs of particles moving
so that one enters apparatus $I$ and the other the apparatus $II$.
There are adjustable apparatus parameters $a$ and $b$. Each apparatus
performs a measurement, giving in output the signals $A(a)$ and
$B(b)$, respectively, where $A,B \in \{-, + \}$.

We distinguish between two cases.

In the $LHV$ case we assume a statistical correlation between the
measurement results due to the information carried by and localized
within each object. This information is part of the content of a set
of hidden variables, denoted collectively by $\lambda$. So, owing to
locality the joint probability on $I$ and $II$ is
\begin{equation}
P(A(a,\lambda),B(b,\lambda)) = P(A(a,\lambda))P(B(b,\lambda))
\end{equation}
mediated over $\lambda$.

In  QM~\cite{grangier} ``one can define {\it entangled} quantum
states of two particles in such a way that their global state is
perfectly defined, whereas the states of the separate particles
remain totally undefined". In this case the joint probability cannot
be factored.

We define the correlation as
\begin{equation}
E(a,b)= P_{+,+}(a,b) + P_{-,-}(a,b) - P_{+,-}(a,b) - P_{-,+}(a,b)
\end{equation}

The aim of the ideal experiment is to test the so called Bell-CHSH
Inequality\cite{4}, stating that the sum
\begin{equation}
S = | E(a,b) - E(a,d)| + | E(c,b) + E(c,d)|
\end{equation}
is $\leq 2$ in $LHV$ case, while  QM predicts a violation of this
inequality till the value $S = 2\sqrt{2}$ is reached.

\subsection{Noise makes QM difficult to support} Each real
measurement apparatus may exhibit a wrong measurement result. We
model this as the classical BSC (Binary Symmetrical
Channel)~\cite{shan}: the signal is the correct one with
probability $1-\varepsilon$ and it is altered by noise with
probability $\varepsilon$, where $0 \leq \varepsilon \leq 1$.

Let $P(A)$ be the probability of $A$ in an ideal apparatus. Given
\begin{equation}
P(A) + P(\overline A) = 1
\end{equation}
where the overline means the opposite result, the probability of $A$
in a noisy apparatus is
\begin{equation}
P_{\varepsilon}(A)= P(A) (1 - \varepsilon) + P(\overline A)
\varepsilon = P(A) (1 - 2 \varepsilon) + \varepsilon
\end{equation}

As also evidenced by Aspect~\cite{aspect} by no means we can consider
the noise on $I$ and $II$ to be equal and independent of the
apparatus parameters. So, in the $LHV$ case the joint probability on
$I$ and $II$ is $$ P_{\varepsilon}(A(a,\lambda),B(b,\lambda)) =
(P(A(a,\lambda)) (1 - 2 \varepsilon_1) + \varepsilon_1))
(P(B(b,\lambda)) (1 - 2 \varepsilon_2) + \varepsilon_2)) = $$ $$ =
P(A(a,\lambda))P(B(b,\lambda)) (1 - 2 \varepsilon_1) (1 - 2
\varepsilon_2) + P(A(a,\lambda)) (1 - 2 \varepsilon_1) \varepsilon_2
$$
\begin{equation}
+ P(B(b,\lambda)) (1 - 2 \varepsilon_2) \varepsilon_1 + \varepsilon_1
\varepsilon_2
\end{equation}
and $$ P_{\varepsilon}(A(a),B(b)) = P(A(a), B(b)) (1 - \varepsilon_1)
(1 - \varepsilon_2) + P({\overline A}(a), B(b)) \varepsilon_1 (1 -
\varepsilon_2) + $$
\begin{equation}
+P(A(a), {\overline B} (b)) (1 - \varepsilon_1) \varepsilon_2 +
P({\overline A}(a), {\overline B} (b)) \varepsilon_1 \varepsilon_2
\end{equation}
in the QM case.

It is easy to verify that in both cases the resulting noisy
correlation is related to the ideal one by \begin{equation}
E_{\varepsilon}(a,b) = (1 - 2 \varepsilon_1) (1 - 2 \varepsilon_2)
E(a,b)
\end{equation}

In this way $S$ becomes $$S_{\varepsilon} = |(1 - 2 \varepsilon_1)
\left[(1 - 2 \varepsilon_2) E(a,b) - (1 - 2 \varepsilon_4) E(a,d)
\right] | + $$
\begin{equation}
+ |(1 - 2 \varepsilon_3) \left[(1 - 2 \varepsilon_2) E(c,b) +
 (1 - 2 \varepsilon_4) E(c,d) \right]|
\end{equation}

The noise parameters have a non trivial role. For example assuming
all such parameters to be equal to $\varepsilon$, $ S_{\varepsilon}$
is certainly lower than 2 $\forall \varepsilon > 0.15$ in both $LHV$
and $QM$ cases.

\subsection{Noise makes LHV difficult to support, too} In testing
Bell's inequality we need to compare the results at the two ends
$I$ and $II$ of the apparatus. Since the measurement apparatus is
a whole, the changing of one parameter has effects on the whole
system. While in most cases this leads to a (3)'s value reduction,
different results can be shown, as we will see in the following
Gedankenexperiment.

Let us consider an example originally given by Bell~\cite{5}. A saw
cuts a coin down the middle, so that the head and the cross are
separated. The two halves (Half Head ($H$) and Half Cross ($C$)) are
sent to the $I$ and $II$ buckets, respectively (see {\bf Figure 1}).

\begin{figure}[htbp]
\centerline{\includegraphics[width=5.02in,height=3.17in]{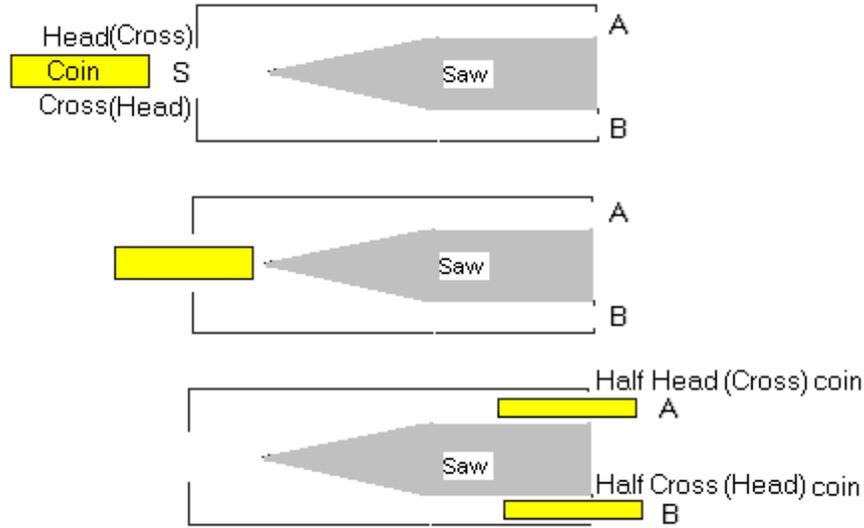}}
\label{fig1} \caption{This figure shows a classical system in
which a saw cuts a coin, so that the head and the cross are
separated and sent into two different buckets. The upper part of
the figure  shows the apparatus and the full coin before entering
the apparatus, the middle part shows the insertion of the coin,
and the lower part the resulting two cut halves of the coin.}
\end{figure}

Over each bucket we place two cameras the first two to detect the
head, the latter two to detect the cross. If there is detection, a
signal is set to on in the circuit of {\bf Figure 2}, otherwise it
remains off. The circuit is made up of four stages $L_1, L_2, L_3,
L_4$ each one including a counter of the concordance between each
couple of incoming signals. Let the camera signals be represented as
$A(a)$, $A(c)$, $B(b)$, and $B(d)$, each of which equals  $+$ or $-$
according with the observation result, and where $a, b, c,$ and $d$
are adjustable parameters.

\begin{figure}[htbp]
\centerline{\includegraphics[width=4.24in,height=1.86in]{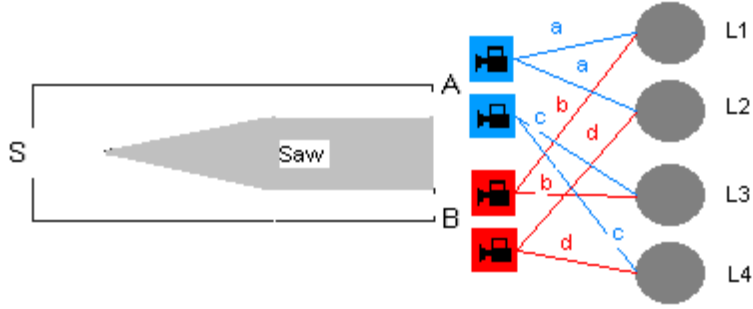}}
\label{fig2} \caption{The figure shows the cutting-coin apparatus,
the four cameras detecting the cross or the head on each half
coin, and the circuit to count the concordance between each couple
of incoming signals.}
\end{figure}

Each stage has two inputs. Each input signal is sent to a stage on a
separate channel subject to noise. The particular conditions and the
choice of $a,b,c,d$ allow potentially a perfect recognition.

The probabilities of the camera signals are as reported in
Table~\ref{tab:tab5}.

\begin{table}[htbp]
\caption{Cameras.}
  \label{tab:tab5}
  \scriptsize
  \begin{center}
    \leavevmode
    \begin{tabular}{|c|c|c|c|c|c|}
      \hline\hline
$A$ & $B$ & $P(A(a),B(b))$ & $P(A(a),B(d))$ & $P(A(c),B(b))$ &
$P(A(c),B(d))$ \\
      \hline\hline
      \hline
+ & + & ${1 \over 2}$& ${1 \over 2}$& ${1 \over 2}$& ${1 \over 2}$\\
+ & - & 0& 0& 0& 0\\ - & + & 0& 0& 0& 0\\ - & - & ${1 \over 2}$& ${1
\over 2}$& ${1 \over 2}$& ${1 \over 2}$\\ \hline\hline
\end{tabular}
  \end{center}
  \end{table}

Now let us assume the channel noises negligible except for $A(a)$ to
$L_2$. For example $A(a)$ may be not well connected to $L_2$ and the
signal falls a certain percentage of times. This kind of error of
misidentification is not so far from the reality of many experiments.
For example Rowe et al.\cite{Rowe} admit that in their experiment a
bright ion is misidentified $ 2 \%$ of the time as being dark.

Let us see the  probabilities and correlation in the counter system,
as reported in Table~\ref{tab:tab5b}.

\begin{table}[htbp]
\caption{ Counter system.}
  \label{tab:tab5b}
  \scriptsize
  \begin{center}
    \leavevmode
    \begin{tabular}{|c|c|c|c|c|c|}
      \hline\hline
$L_i()$ & $L_i()$ & $P_{\varepsilon}(L_1(a), L_1(b))$ &
$P_{\varepsilon} (L_2(a), L_2(d))$ & $P_{\varepsilon} (L_3(c),
L_3(b))$ & $P_{\varepsilon}(L_4(c), L_4(d))$ \\
      \hline\hline
      \hline
+ & + & ${1 \over 2}$   & ${1 \over 2} (1-\varepsilon)$ & ${1 \over
2}$& ${1 \over 2}$\\ + & - & 0       & 0 & 0 & 0\\ - & + & 0       &
${1 \over 2} \varepsilon$ & 0 & 0\\ - & - & ${1 \over 2}$   & ${1
\over 2}$& ${1 \over 2}$& ${1 \over 2}$\\ \hline
  &
& $E_{\varepsilon} (a,b) = 1$ & $E_{\varepsilon} (a,d) =
1-\varepsilon$ & $E_{\varepsilon} (c,b) = 1$ & $E_{\varepsilon} (c,d)
= 1$ \\ \hline\hline
\end{tabular}
  \end{center}
  \end{table}

It is easy to verify that the Einstein-Bell locality has been
preserved, but, by collecting the results in tables we have
\begin{equation}
| E_{\varepsilon} (a,b) - E_{\varepsilon} (a,d)| + | E_{\varepsilon}
(c,b) + E_{\varepsilon} (c,d)| = 2 + \varepsilon
\end{equation}
The result is that a macroscopic system can violate the Bell's
inequality $\forall \varepsilon>0$. In the light of this example the
violation of Bell's inequality could be a good test to verify the
existence of faulty instruments in a classical complex system.

\subsection{Low Detection Efficiency may lead to unrepresentative
results} Between the known loopholes in Bell's Inequality testing,
the problem of the detection efficiency was identified as long ago
as 1970, and it is also called the Enhancement Loophole. It arises
because only a small subset of the pairs emitted are actually
detected, so certain highly model-dependent assumptions have to be
made about the statistical sampling\cite{[66]}\cite{cla}.

We can model this fact by erasure channel~\cite{shan}: the particle
is detected with probability $1-\delta$ and it is not detected by
with probability $\delta$, where $0 \leq \delta \leq 1$. So the
probability of simultaneous detection is  the product of the
detection probability of each apparatus $I$ and $II$, and only a
fraction $(1-\delta(A(a)))(1-\delta(B(b)))$ of the correlated pairs
is actually detected. When $\delta$'s parameters grow we would have
to assume that the sample of pairs registered is not a faithful
representative of the whole ensemble emitted.

For example let us consider a sample of $N$ particle pairs belonging
to two distinct classes $C_1$ and $C_2$ such that $E_{C_1}(a,b)=1$,
$E_{C_2}(a,b)=-1$, and $\#C_1 = \#C_2$. Then $E(a,b)=0$. Of the $N$
couples we detect only a fraction $\phi$. Then the probability of an
equilibrate sampling between the two classes is just $$\left({{N/2}
\atop {N\phi/2}}\right) \left({{N/2} \atop {N\phi/2}}\right) \left(
{1\over 2} \right)^{N\phi} $$

\bigskip

This corresponds to the introduction of free parameters $\Delta$ in
$S$, leading to
\begin{equation}
S_{\delta} = | \Delta_1 E(a,b)
-
\Delta_2 E(a,d) | + | \Delta_3 E(c,b) + \Delta_4 E(c,d)|
\end{equation}
where $-{1 \over {E(a_i,b_i)} } \leq \Delta_i \leq {1 \over
{E(a_i,b_i)} }$

So we agree with Weihs et al.~\cite{viola} that ``an ultimate
experiment should also have higher detection/collection efficiency,
of the common $5\%$ of actual experiments", and that this loophole
makes still possible ``unlikely, local realistic or semi­classical
interpretations".

\subsection{Ideal Detection Efficiency may lead to unrepresentative
results, too} Even in the case of a perfect efficiency of the
detectors (the Rowe's experiment is the first to  point in this
direction),  some problems might occur with the sample of N
particles pairs used during the test.

Let us consider a Rowe et al. like experiment~\cite{Rowe}, involving
four sets of phase angles chosen to apply the CHSH inequality. If we
assume that hidden variables are in fact involved and the hidden
variables of the $N_1$ pairs of the first set of phases are different
from the $N_2$ of the second set, then the CHSH inequality becomes $
S \leq 4$~\cite{adenier} and we cannot discriminate between QM and
LHV.

In order to avoid this problem, we can calculate the probability to
have the values $ \lambda_1, \lambda_2, \ldots, \lambda_n $ of the
first set $N_1$ equal to the $\lambda_1, \lambda_2, \ldots,
\lambda_n$ of $N_2$. For example if $N_1 = N_2= n$ and if $\lambda$
can assume a discrete number of values $N_{TOT}$ in the interval $0
\leq \lambda \leq 2\pi$ with an isotropic distribution (in the
continuum case we would have $\rho(\lambda) = 1/2\pi$), we obtain
\begin{equation}
P = \prod_{i=0}^{n-1} \frac{n-i}{N_{TOT}-i}
\end{equation}

In the case of Rowe et al. experiment $n=20000$ but nobody knows
$N_{TOT}$. If $N_{TOT} \rightarrow \infty$ we have $ P \rightarrow 0$
and the experiment cannot discriminate between QM and LHV.

\section{Conclusions}
Bell's inequality tests necessitate major improvements of technology
in order to finally, after more than 15 years, go significantly
beyond the 1982 experiment of Aspect et al.~\cite{a2}. While
expecting that any improved experiment will also agree with quantum
theory, actually the final answer to the eternal question: ``Is the
moon there, when nobody looks?'', is certainly up to our judgement
capability. But sometime also the question "Is the moon there when we
look at it by a noisy telescope?" appears very hard to address.

\end{document}